\documentclass[twocolumn,twoside,slac_two]{revtex4}
\usepackage{graphicx}
\usepackage{fancyhdr}
\begin{document}

\title{Discovery of a Young Gamma-ray Pulsar Associated with an Extended TeV Gamma-ray Source}

%

\author{M. Dormody on behalf of the Fermi-LAT collaboration}
\affiliation{UCSC, Santa Cruz, CA 95060, USA}

\begin{abstract}
Since its launch in June 2008, the Large Area Telescope (LAT), onboard
the \emph{Fermi} Gamma-ray Space Telescope, has greatly added to our
understanding of gamma-ray pulsars. Its fine point spread function and
large effective area, combined with the time-differencing method, make
it the first gamma-ray instrument capable of discovering a new
population of gamma-ray pulsars. We will present the recent discovery
of the youngest ($\tau\sim4600$ yr) radio-quiet gamma-ray pulsar
discovered in a blind frequency search so far: PSR J1022-5746, a
pulsar associated with an extended TeV source. We also present
multiwavelength observations of the source, including X-ray
observations.

\end{abstract}

\maketitle

\thispagestyle{fancy}


\section{Overview}

There is an interesting cluster of bright gamma-ray sources located
within the Galactic plane, roughly at $l \sim 285^\circ$, extending
about $\pm 5^{\circ}$ . This region houses six Fermi sources, most of
them gamma-ray pulsars. These sources include radio-loud pulsars PSR
J1048-5832 and PSR J1028-5819 and radio-quiet PSR J1044-5737, recently
discovered in a blind frequency search \citep{8new}. The new pulsar,
PSR J1022-5746, is located within this region.

This source was found early on in the mission as a bright gamma-ray
source, published in the Fermi Bright Source List (0FGL J1024.0-5754)
\citep{BrightSourceList}. However, it was not detected in a blind
frequency search for several reasons. The 0FGL position is $\sim 17'$
away from the timed pulsar location, which is insufficient for blind
pulsation searches. In addition, this pulsar has a non-neglible $\ddot
f$ correction. These two factors made detection at an early date very
challenging.

We will discuss how J1022-5746 was discovered in a blind search, list
its timing properties, and discuss multiwavelength observations,
especially in the TeV regeme.

\section{Detection in a Blind Frequency Search}

Gamma-ray photon data from pulsars is very sparse, requiring long
integration times for sufficient statistics. The length of the
integration time, or viewing period, increases the resolution of the
FFT used to detect the periodic signal since the number of FFT bins
$N$ is $N = 2 T f_{\rm max}$. Moreover, the long viewing period
requires us to scan tens of thousands of $\dot f$ trials to correct
for the spin-down of the pulsar. In addition, young pulsars typically
have timing noise, or can suddenly and abruptly change their
rotational frequencies (known as glitches), making an FFT of the full
time series computationally intensive.

Instead, we look for periodicity using the \emph{time-differencing}
method \citep{atwood06}. Since periodicity in photon arrival times
will also be present in the differences between arrival times, we
calculate all differences between arrival times up to a maximum
differencing window of about 6 days ($T = 2^{19}$ s). The loss in
sensitivity is made up for by the reduced number of $\dot f$ trials,
and the computational time is dramatically reduced. We cover zero
spin-down to the spin-down of the Crab pulsar ($\frac{\dot f}{f} =
-1.125 \times 10^{-11}$ s$^{-1}$). This method is very efficient at
finding pulsars and has resulted in the discovery of 16 pulsars in the
first few months of the mission \citep{LATBSPs}, including PSR
J1022-5746.

PSR J1022-5746 has the largest $\dot f$ and $\ddot f$ of all the
pulsars found in the blind search, making it difficult to find in a
long span of data without correcting for $\ddot f$.

We used a circular region of interest (ROI) of radius $R \le
0.8^{\circ}$, minimum energy of 300 MeV, and \emph{diffuse} class
photons \citep{LATCalib}.

\section{Pulsar Properties}

\begin{figure*}[htp]
\centering
\includegraphics[width=4in]{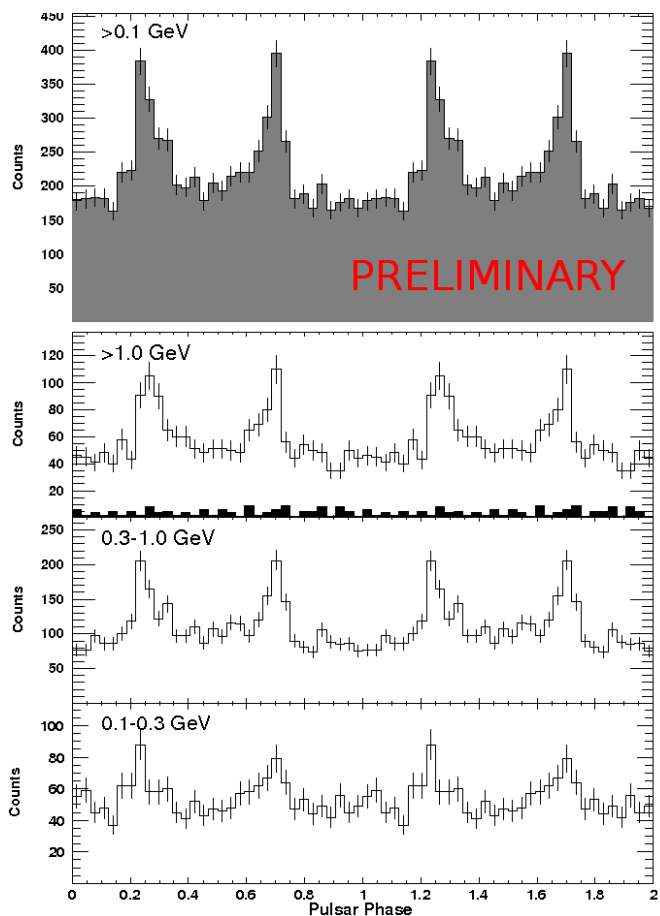}
\caption{Double-phase light curve of PSR J1022-5746 across five energy
  bands: $>$ 100 MeV, 100-300 MeV, 300 MeV-1 GeV, $>$ 1 GeV, and $>$ 5
  GeV.} \label{1022_banded}
\end{figure*}

The double-phase light curve of the pulsar is found in figure
\ref{1022_banded}. We can see the double peak structure across the
multiple energy bands ($>$ 100 MeV, 100-300 MeV, 300 MeV-1 GeV, $>$ 1
GeV, $>$ 5 GeV). Also, we have shifted the first peak to $\phi = 0.25$
for clarity. The peak separation is $0.45 \pm 0.01$ and the off-pulse
region is $0.75-1.18$.

\begin{figure*}[htp]
\centering
\includegraphics[width=5in]{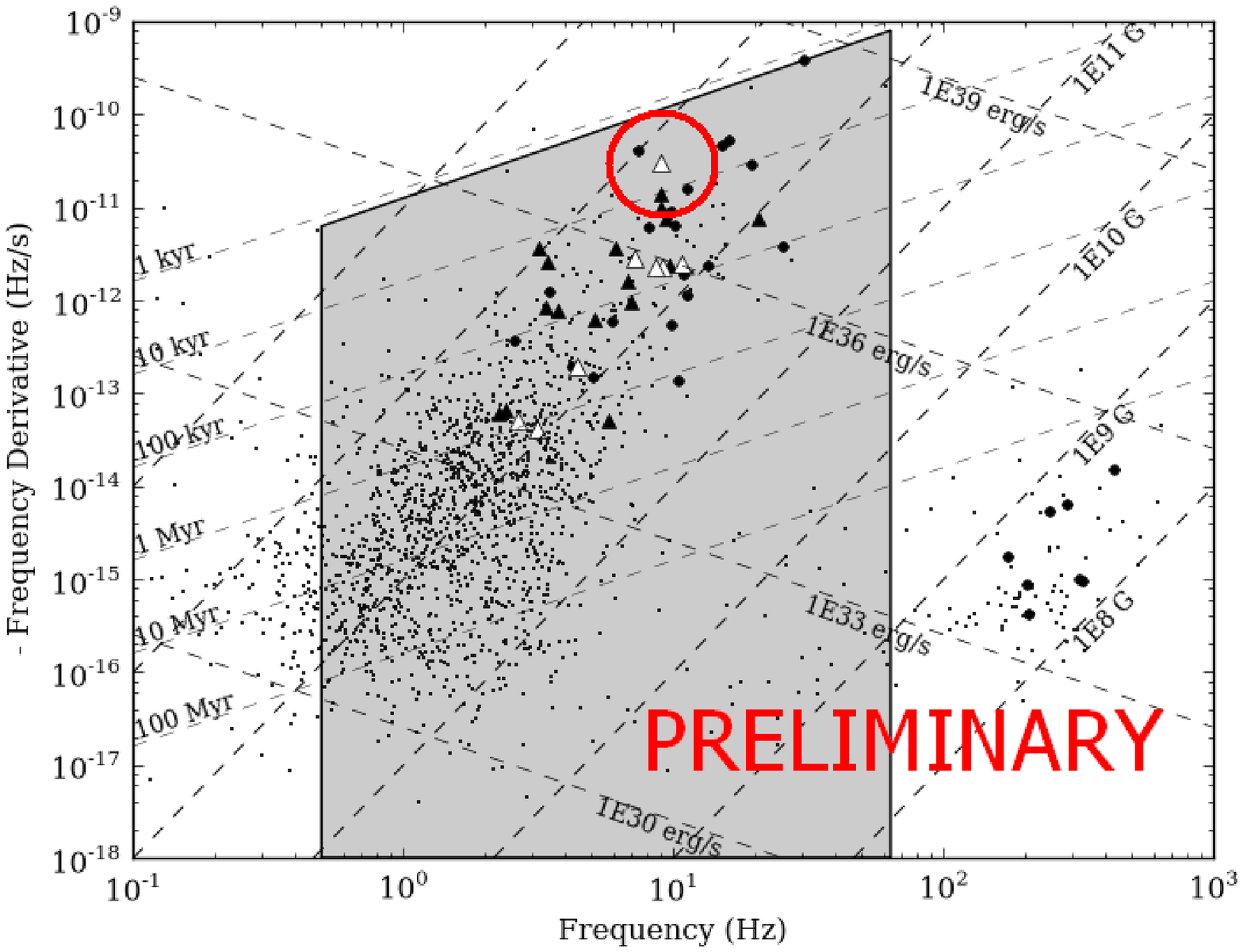}
\caption{Parameter space of the known pulsars. The pulsars in the ATNF
  database are indicated by black dots, and the radio-selected
  $\gamma$-ray pulsars are indicated by black circles. The
  $\gamma$-ray selected pulsars are denoted with solid triangles for
  previously reported pulsars. PSR J1022-5746, along with seven other
  pulsars \citep{8new} are denoted with unfilled triangles. PSR
  J1022-5746 is represented as the unfilled triangle located inside
  the red circle, indicating its position with respect to the other
  gamma-ray pulsars.} \label{f0_v_f1_circle}
\end{figure*}

We can compare this pulsar to the known pulsar population in figure
\ref{f0_v_f1_circle}. This pulsar is one of the youngest gamma-ray
pulsars, and the youngest pulsar discovered in a blind search. Also,
it is the most highly energetic blind search pulsar. The rotational
ephemerides can be found in table \ref{timing_solutions}.

\begin{table}[htp]
\begin{center}
\caption{Names and locations of the new gamma-ray pulsars.}
\begin{tabular}{|l|l|}
\hline
\textbf{Pulsar Name} & J1022--5746 \\
\textbf{Source Associations} & 0FGL J1024.0--5754 \\
& HESS J1023--575 \\
&  CXOU J102302.8--574607 \\
\textbf{R.A.} & 155.7597 \\
\textbf{Decl.} &  -57.7693 \\
\textbf{$l$} & 284.2 \\
\textbf{$b$} & -0.4 \\
\hline
\end{tabular}
\label{pulsar_properties}
\end{center}
\end{table}

\begin{table}[htp]
\begin{center}
\caption{Rotational ephemerides for the new pulsars. For all timing
  solutions, the reference epoch is MJD 54800. Row 1 gives the
  pulsar name. Row 2 lists the number of photons from the standard
  cut over the 11 month observational period. Row 3 gives the flux
  above 100 MeV in units of $10^{-8}$ cm$^{-2}$s$^{-1}$ using P6v3
  IRFs. Rows 4 and 5 list the frequency in units of Hz and
  frequency derivative in units of $-10^{-12}$ Hz s$^{-1}$. Rows 6,
  7 and 8 give the characteristic age in units of kyr, spin-down
  luminosity in units of $10^{34}$ erg s$^{-1}$, and magnetic field
  strength at the light cylinder $B_{LC}$ determined from the spin
  parameters. These derived parameters are rounded to the nearest
  significant digit.}
\begin{tabular}{|l|r|}
\hline
\textbf{Pulsar Name} & J1022--5746 \\
\textbf{$n_{\gamma}$} & 4365 \\
\textbf{$F_{100}$ ($10^{-8}$ cm$^{-2}$ s$^{-1}$)} & $23.0 \pm 2.6$ \\
\textbf{$f$ (Hz)} & 8.970977214(3) \\
\textbf{$\dot{f}$ ($-10^{-12}$ Hz s$^{-1}$)} & 30.9163(3) \\
\textbf{$\ddot{f}$ ($10^{-21}$ Hz s$^{-2}$)} & 6.49(9) \\
\textbf{$\tau$ (kyr)} & 4.6 \\
\textbf{$\dot{E}$ ($10^{34}$ erg s$^{-1}$)} & 1094.9 \\
\textbf{$B_{LC}$ (kG)} &  44.0 \\
\hline
\end{tabular}
\label{timing_solutions}
\end{center}
\end{table}

\section{Multiwavelength}

\subsection{TeV}

This pulsar might help to explain the engine powering the TeV source
HESS J1023-575. HESS reported detection of an extended TeV source near
the Westerlund 2 star cluster \citep{2007A&A...467.1075A}. The
possible VHE emission explanations put forth were a massive WR binary
system WR 20a, a young stellar cluster Westerlund 2, or cosmic rays
accelerated at their termination shock and interacting with their
environment.

\begin{figure*}[htp]
\centering
\includegraphics[width=3.5in]{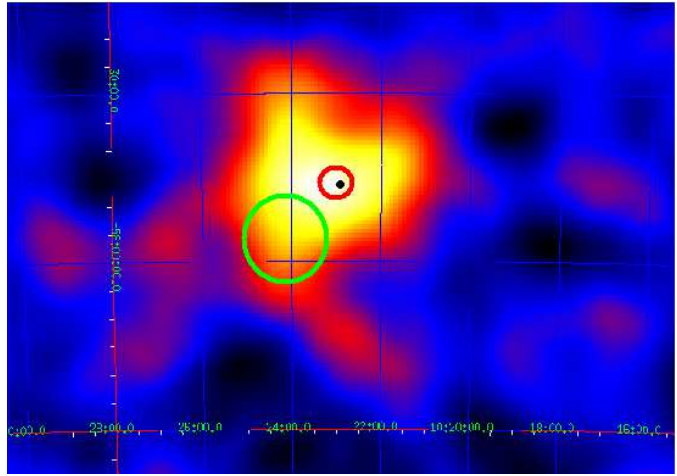}
\caption{Map of the HESS image J1023-575
  \citep{2007A&A...467.1075A}. The green circle shows Bright Source
  List location, the red circle shows latest LAT source location, and
  the black dot shows the timed pulsar location.} \label{HESS_image}
\end{figure*}

When compared with the location of the pulsar J1022-5746, it is
apparent that the pulsar lies close to the central region of the TeV
source. The bright source location lies $18'$ away from the LAT
catalog location, and the timed pulsar location lies nicely within the
TeV source, as seen in figure \ref{HESS_image}.

\subsection{Infrared}

\begin{figure*}[htp]
\centering
\includegraphics[width=3in]{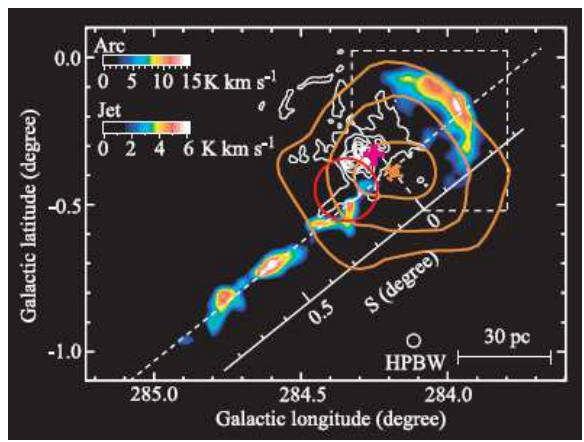}
\caption{Distribution of $^{12}$CO emission for the arc and jet
  \citep{2009PASJ...61L..23F}. The pulsar source lies near the center
  of HESS source (orange cross), and the Westerlund 2 cluster is
  located at the pink cross.} \label{fukui_image}
\end{figure*}

Fukui et. al. 2009 observed a jet and arc of molecular gas aligning
with this HESS source, possibly caused by an anisotropic supernova
explosion, as seen in figure \ref{fukui_image}
\citep{2009PASJ...61L..23F}.

\subsection{X-ray}

A \emph{Chandra} 130 ks image reveals a faint source named CXOU
J102302.8-574607 as the likely counterpart. This is coincident with
pulsar location to within $0.1'$. The column density $N_H \sim 1.3
\times 10^{22}$ cm$^{-2}$ implies $d > 10$ kpc. J1022-5746 is located
$> 8'$ away from the Westerlund 2 core ($\sim 8$ kpc).

\section{Conclusions}

PSR J1022-5746 is the youngest and most energetic gamma-selected
gamma-ray pulsar discovered. It has been observed in radio, but no
pulsations were detected. It was discovered in recent blind frequency
searches, along with 7 additional pulsars. A Chandra analysis reveals
faint, distant X-ray source, about $8'$ away from the core of the
Westerlund 2 cluster. PSR J1022-5746 is coincident with TeV source
HESS J1023-575, suggesting the pulsar contributes to the VHE emission.

\bigskip 
\begin{acknowledgments}

The \emph{Fermi} LAT Collaboration acknowledges support from a number
of agencies and institutes for both development and the operation of
the LAT as well as scientific data analysis. These include NASA and
DOE in the United States, CEA/Irfu and IN2P3/CNRS in France, ASI and
INFN in Italy, MEXT, KEK, and JAXA in Japan, and the K.~A.~Wallenberg
Foundation, the Swedish Research Council and the National Space Board
in Sweden. Additional support from INAF in Italy and CNES in France
for science analysis during the operations phase is also gratefully
acknowledged.  Much of the work presented here was carried out on the
UCSC Astronomy department's Pleiades supercomputer. This work made
extensive use of the ATNF pulsar catalogue. We thank N. Gehrels and
the rest of the \textit{Swift} team for the \textit{Swift}/XRT
observations of the LAT error circles of several of these
newly-discovered pulsars.

\end{acknowledgments}

\bigskip 

\bibliography{journapj,Fermi_Bibtex_full_v2,my_bibliography,extra_bibliography}

\begin{thebibliography}{7}
\expandafter\ifx\csname natexlab\endcsname\relax\def\natexlab#1{#1}\fi
\expandafter\ifx\csname bibnamefont\endcsname\relax
  \def\bibnamefont#1{#1}\fi
\expandafter\ifx\csname bibfnamefont\endcsname\relax
  \def\bibfnamefont#1{#1}\fi
\expandafter\ifx\csname citenamefont\endcsname\relax
  \def\citenamefont#1{#1}\fi
\expandafter\ifx\csname url\endcsname\relax
  \def\url#1{\texttt{#1}}\fi
\expandafter\ifx\csname urlprefix\endcsname\relax\def\urlprefix{URL }\fi
\providecommand{\bibinfo}[2]{#2}
\providecommand{\eprint}[2][]{\url{#2}}

\bibitem[{\citenamefont{{Abdo} et~al.}(2009{\natexlab{a}})\citenamefont{{Abdo},
  {Ackermann}, {Atwood}, {Baldini}, {Ballet}, {Barbiellini}, {Baring},
  {Bastieri}, and {others}}}]{8new}
\bibinfo{author}{\bibfnamefont{A.~A.} \bibnamefont{{Abdo}}},
  \bibinfo{author}{\bibfnamefont{M.}~\bibnamefont{{Ackermann}}},
  \bibinfo{author}{\bibfnamefont{W.~B.} \bibnamefont{{Atwood}}},
  \bibinfo{author}{\bibfnamefont{L.}~\bibnamefont{{Baldini}}},
  \bibinfo{author}{\bibfnamefont{J.}~\bibnamefont{{Ballet}}},
  \bibinfo{author}{\bibfnamefont{G.}~\bibnamefont{{Barbiellini}}},
  \bibinfo{author}{\bibfnamefont{M.~G.} \bibnamefont{{Baring}}},
  \bibinfo{author}{\bibfnamefont{D.}~\bibnamefont{{Bastieri}}},
  \bibnamefont{and} \bibinfo{author}{\bibnamefont{{others}}}
  (\bibinfo{year}{2009}{\natexlab{a}}), \bibinfo{note}{in preparation}.

\bibitem[{\citenamefont{{Abdo}}(2009)}]{BrightSourceList}
\bibinfo{author}{\bibfnamefont{A.~A.} \bibnamefont{{Abdo}}},
  \bibinfo{journal}{ArXiv e-prints}  (\bibinfo{year}{2009}),
  \bibinfo{note}{\apjs , submitted - Bright Source List paper},
  \eprint{0902.1340}.

\bibitem[{\citenamefont{{Atwood} et~al.}(2006)\citenamefont{{Atwood},
  {Ziegler}, {Johnson}, and {Baughman}}}]{atwood06}
\bibinfo{author}{\bibfnamefont{W.~B.} \bibnamefont{{Atwood}}},
  \bibinfo{author}{\bibfnamefont{M.}~\bibnamefont{{Ziegler}}},
  \bibinfo{author}{\bibfnamefont{R.~P.} \bibnamefont{{Johnson}}},
  \bibnamefont{and} \bibinfo{author}{\bibfnamefont{B.~M.}
  \bibnamefont{{Baughman}}}, \bibinfo{journal}{\apjl}
  \textbf{\bibinfo{volume}{652}}, \bibinfo{pages}{L49} (\bibinfo{year}{2006}).

\bibitem[{\citenamefont{{Abdo} et~al.}(2009{\natexlab{b}})\citenamefont{{Abdo},
  {Ackermann}, {Ajello}, {Anderson}, {Atwood}, {Axelsson}, {Baldini}, {Ballet},
  {Barbiellini}, {Baring} et~al.}}]{LATBSPs}
\bibinfo{author}{\bibfnamefont{A.~A.} \bibnamefont{{Abdo}}},
  \bibinfo{author}{\bibfnamefont{M.}~\bibnamefont{{Ackermann}}},
  \bibinfo{author}{\bibfnamefont{M.}~\bibnamefont{{Ajello}}},
  \bibinfo{author}{\bibfnamefont{B.}~\bibnamefont{{Anderson}}},
  \bibinfo{author}{\bibfnamefont{W.~B.} \bibnamefont{{Atwood}}},
  \bibinfo{author}{\bibfnamefont{M.}~\bibnamefont{{Axelsson}}},
  \bibinfo{author}{\bibfnamefont{L.}~\bibnamefont{{Baldini}}},
  \bibinfo{author}{\bibfnamefont{J.}~\bibnamefont{{Ballet}}},
  \bibinfo{author}{\bibfnamefont{G.}~\bibnamefont{{Barbiellini}}},
  \bibinfo{author}{\bibfnamefont{M.~G.} \bibnamefont{{Baring}}},
  \bibnamefont{et~al.}, \bibinfo{journal}{Science}
  \textbf{\bibinfo{volume}{325}}, \bibinfo{pages}{840}
  (\bibinfo{year}{2009}{\natexlab{b}}).

\bibitem[{\citenamefont{{Abdo} et~al.}(2009{\natexlab{c}})\citenamefont{{Abdo},
  {Ackermann}, {Atwood}, {Baldini}, {Ballet}, {Barbiellini}, {Baring},
  {Bastieri}, and {others}}}]{LATCalib}
\bibinfo{author}{\bibfnamefont{A.~A.} \bibnamefont{{Abdo}}},
  \bibinfo{author}{\bibfnamefont{M.}~\bibnamefont{{Ackermann}}},
  \bibinfo{author}{\bibfnamefont{W.~B.} \bibnamefont{{Atwood}}},
  \bibinfo{author}{\bibfnamefont{L.}~\bibnamefont{{Baldini}}},
  \bibinfo{author}{\bibfnamefont{J.}~\bibnamefont{{Ballet}}},
  \bibinfo{author}{\bibfnamefont{G.}~\bibnamefont{{Barbiellini}}},
  \bibinfo{author}{\bibfnamefont{M.~G.} \bibnamefont{{Baring}}},
  \bibinfo{author}{\bibfnamefont{D.}~\bibnamefont{{Bastieri}}},
  \bibnamefont{and} \bibinfo{author}{\bibnamefont{{others}}},
  \bibinfo{journal}{\apj}  (\bibinfo{year}{2009}{\natexlab{c}}),
  \eprint{0904.2226}.

\bibitem[{\citenamefont{{Aharonian} et~al.}(2007)\citenamefont{{Aharonian},
  {Akhperjanian}, {Bazer-Bachi}, {Beilicke}, {Benbow}, {Berge}, {Bernl{\"o}hr},
  {Boisson}, {Bolz}, {Borrel} et~al.}}]{2007A&A...467.1075A}
\bibinfo{author}{\bibfnamefont{F.}~\bibnamefont{{Aharonian}}},
  \bibinfo{author}{\bibfnamefont{A.~G.} \bibnamefont{{Akhperjanian}}},
  \bibinfo{author}{\bibfnamefont{A.~R.} \bibnamefont{{Bazer-Bachi}}},
  \bibinfo{author}{\bibfnamefont{M.}~\bibnamefont{{Beilicke}}},
  \bibinfo{author}{\bibfnamefont{W.}~\bibnamefont{{Benbow}}},
  \bibinfo{author}{\bibfnamefont{D.}~\bibnamefont{{Berge}}},
  \bibinfo{author}{\bibfnamefont{K.}~\bibnamefont{{Bernl{\"o}hr}}},
  \bibinfo{author}{\bibfnamefont{C.}~\bibnamefont{{Boisson}}},
  \bibinfo{author}{\bibfnamefont{O.}~\bibnamefont{{Bolz}}},
  \bibinfo{author}{\bibfnamefont{V.}~\bibnamefont{{Borrel}}},
  \bibnamefont{et~al.}, \bibinfo{journal}{\aap} \textbf{\bibinfo{volume}{467}},
  \bibinfo{pages}{1075} (\bibinfo{year}{2007}),
  \eprint{arXiv:astro-ph/0703427}.

\bibitem[{\citenamefont{{Fukui} et~al.}(2009)\citenamefont{{Fukui}, {Furukawa},
  {Dame}, {Dawson}, {Yamamoto}, {Rowell}, {Aharonian}, {Hofmann}, {de O{\~n}a
  Wilhelmi}, {Minamidani} et~al.}}]{2009PASJ...61L..23F}
\bibinfo{author}{\bibfnamefont{Y.}~\bibnamefont{{Fukui}}},
  \bibinfo{author}{\bibfnamefont{N.}~\bibnamefont{{Furukawa}}},
  \bibinfo{author}{\bibfnamefont{T.~M.} \bibnamefont{{Dame}}},
  \bibinfo{author}{\bibfnamefont{J.~R.} \bibnamefont{{Dawson}}},
  \bibinfo{author}{\bibfnamefont{H.}~\bibnamefont{{Yamamoto}}},
  \bibinfo{author}{\bibfnamefont{G.~P.} \bibnamefont{{Rowell}}},
  \bibinfo{author}{\bibfnamefont{F.}~\bibnamefont{{Aharonian}}},
  \bibinfo{author}{\bibfnamefont{W.}~\bibnamefont{{Hofmann}}},
  \bibinfo{author}{\bibfnamefont{E.}~\bibnamefont{{de O{\~n}a Wilhelmi}}},
  \bibinfo{author}{\bibfnamefont{T.}~\bibnamefont{{Minamidani}}},
  \bibnamefont{et~al.}, \bibinfo{journal}{\pasj} \textbf{\bibinfo{volume}{61}},
  \bibinfo{pages}{L23+} (\bibinfo{year}{2009}), \eprint{0903.5340}.

\end{thebibliography}

\end{document}